\documentclass{moriond}

\usepackage{amssymb}
\usepackage{xspace}

\def\be{\begin{equation}}
\def\ee{\end{equation}}
\def\bea{\begin{eqnarray}}
\def\eea{\end{eqnarray}}

\newcommand{\alphas}{\alpha_{\rm s}}
\newcommand{\alphasmZ}{\alphas(\rm m^2_{_{\rm Z}})}
\newcommand{\sqrts}{\sqrt{\rm s}}

\newcommand{\epem}{e^+e^-}

\providecommand{\bbbar}{\rm b\overline{b}}
\providecommand{\ccbar}{c\overline{c}}
\providecommand{\ttbar}{t\overline{t}}

\newcommand{\Rlz}   {R^0_\ell}
\newcommand{\Ghad}  {\rm\Gamma_{\mathrm{had}}}

\newcommand{\GZ}    {\rm \Gamma_{\mathrm{Z}}}
\newcommand {\so}   {\rm \sigma_0^{had}}

\newcommand*{\eg}{e.g.\@\xspace}
\newcommand*{\ie}{i.e.\@\xspace}
\def\ttt#1{\texttt{\small #1}}

\newcommand{\Photo}{\includegraphics[height=30mm]{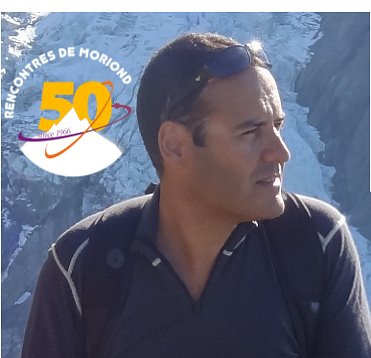}}

\begin{document}
\title{\huge $\alphas$ review (2016) \vspace*{-0.3cm}}

\author{\underline{David d'Enterria}$^1$}

\address{$^1$  CERN, EP Department, CH-1211 Geneva 23, Switzerland}

\maketitle\abstracts{
The current world-average of the strong coupling at the Z pole mass, 
$\alphasmZ = 0.1181~\pm~0.0013$, is obtained from a comparison of perturbative 
QCD calculations computed, at least, at next-to-next-to-leading-order accuracy, to a set of 
6 groups of experimental observables: (i) lattice QCD ``data'', (ii) $\tau$ hadronic decays, 
(iii)~proton structure functions, 
(iv)~event shapes and jet rates in $\epem$ collisions, 
(v)~Z boson hadronic decays, and (vi)~top-quark cross sections in p-p collisions. 
In addition, at least 8 other $\alphas$ extractions, 
usually with a lower level of theoretical and/or experimental accuracy today, have been 
proposed: pion, $\Upsilon$, W hadronic decays; soft and hard fragmentation functions; jets 
cross sections in pp, e-p and $\gamma$-p collisions; and photon F$_2$ structure function in 
$\gamma\,\gamma$ collisions. These 14 $\alphas$ determinations are reviewed, and the
perspectives of reduction of their present uncertainties are discussed.
}

\section{Introduction}

The strong coupling $\alphas$, one of the fundamental parameters of the Standard Model, sets the
scale of the strength of the strong interaction between quarks and gluons, theoretically described by 
Quantum Chromodynamics (QCD)~\cite{d'Enterria:2015toz}. 
Its current value at the reference Z pole mass amounts~\cite{PDG} to $\alphasmZ$~=~0.1186~$\pm$~0.0013, 
with a $\delta\alphasmZ/\alphasmZ \approx$~1\% uncertainty---orders of magnitude larger than that of the
gravitational ($\rm \delta G/G\approx 10^{-5}$), Fermi ($\rm \delta G_{\rm F}/G_{\rm F}\approx
10^{-8}$), and QED ($\rm \delta \alpha/\alpha\approx 10^{-10}$) couplings, making of $\alphas$ the least 
precisely known of all fundamental constants in nature. Improving our knowledge of $\alphas$ is a 
prerequisite to reduce the theoretical uncertainties in the calculations of all high-precision 
perturbative QCD (pQCD) observables whose cross sections or decay rates depend on higher-order powers 
of $\alphas$, as is the case for virtually all those accessible at the LHC. Chiefly, in the Higgs sector, the 
$\alphas$ uncertainty is currently the second major contributor (after the bottom mass) to the parametric 
uncertainties of its dominant $\rm H\to\bbbar$ partial decay, and it's the leading one for the 
$\rm H\to\ccbar,g\,g$ branching fractions. The $\alphas$ running impacts also our understanding 
of physics approaching the Planck scale, \eg\ the stability of the electroweak vacuum~\cite{Buttazzo:2013uya} 
or the scale at which the interaction couplings unify.\\

The latest update of the Particle-Data-Group (PDG) world-average $\alphasmZ$, obtained from a comparison of 
next-to-next-to-leading-order (NNLO) pQCD calculations to a set of 6 groups of experimental observables, has resulted 
in a factor of two increase in the $\alphas$ uncertainty, compared to the previous (2014) PDG value~\cite{PDG}. 
This fact calls for new independent approaches to determine $\alphas$ from the data, with experimental and
theoretical uncertainties different from those of the methods currently used, in order to reduce 
the overall uncertainty of the $\alphas$ world-average. 
These proceedings provide a summary of all the $\alphas$ determination methods described in detail in refs.~\cite{d'Enterria:2015toz,PDG} 
where more complete lists of references can be found.

\section{Current world $\alphasmZ$ average}
\label{sec:current}

\begin{figure}[htpb!]
\centerline{
\includegraphics[width=0.40\linewidth,height=10.75cm]{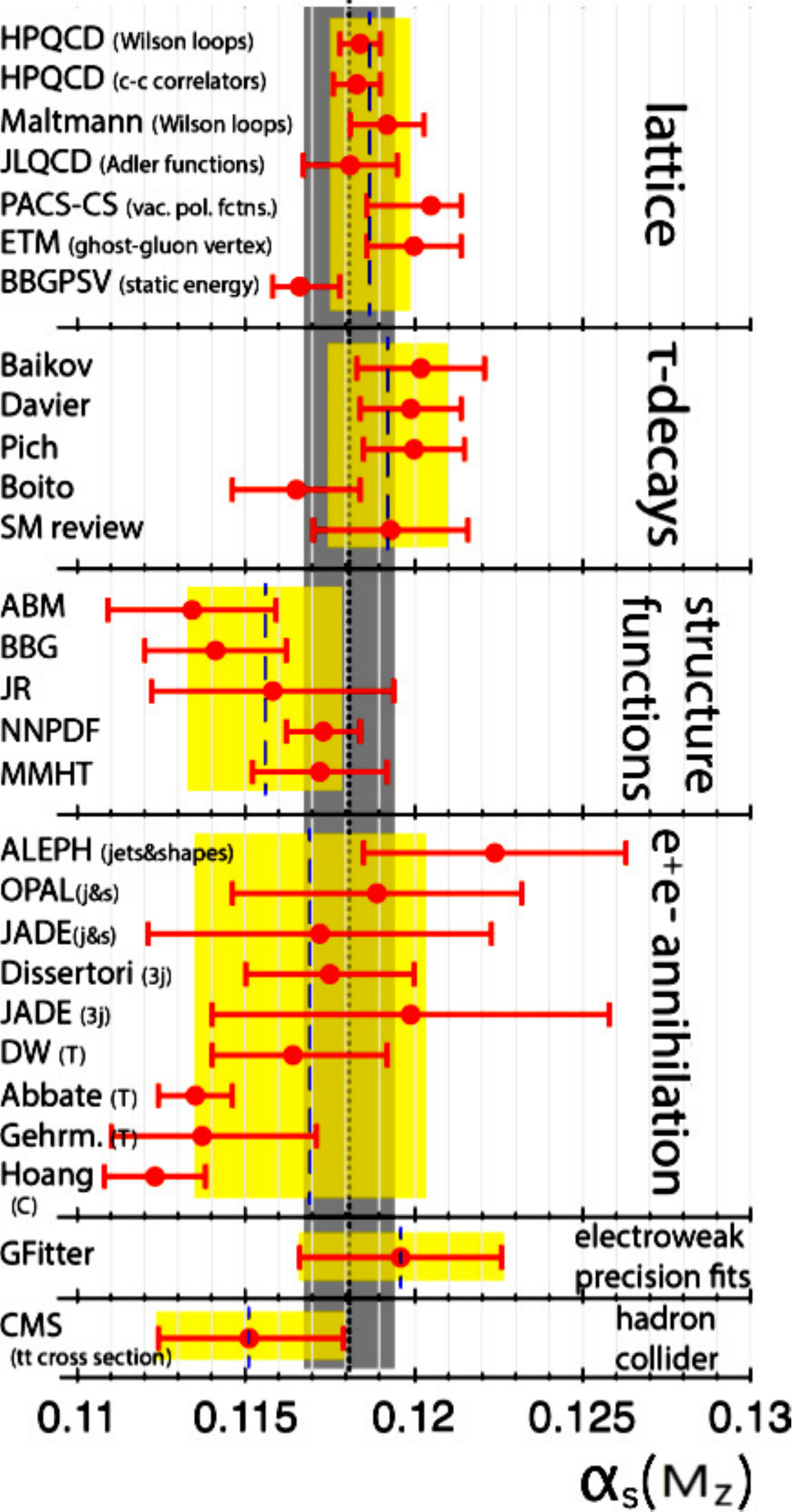}
\includegraphics[width=0.60\linewidth,height=10.5cm]{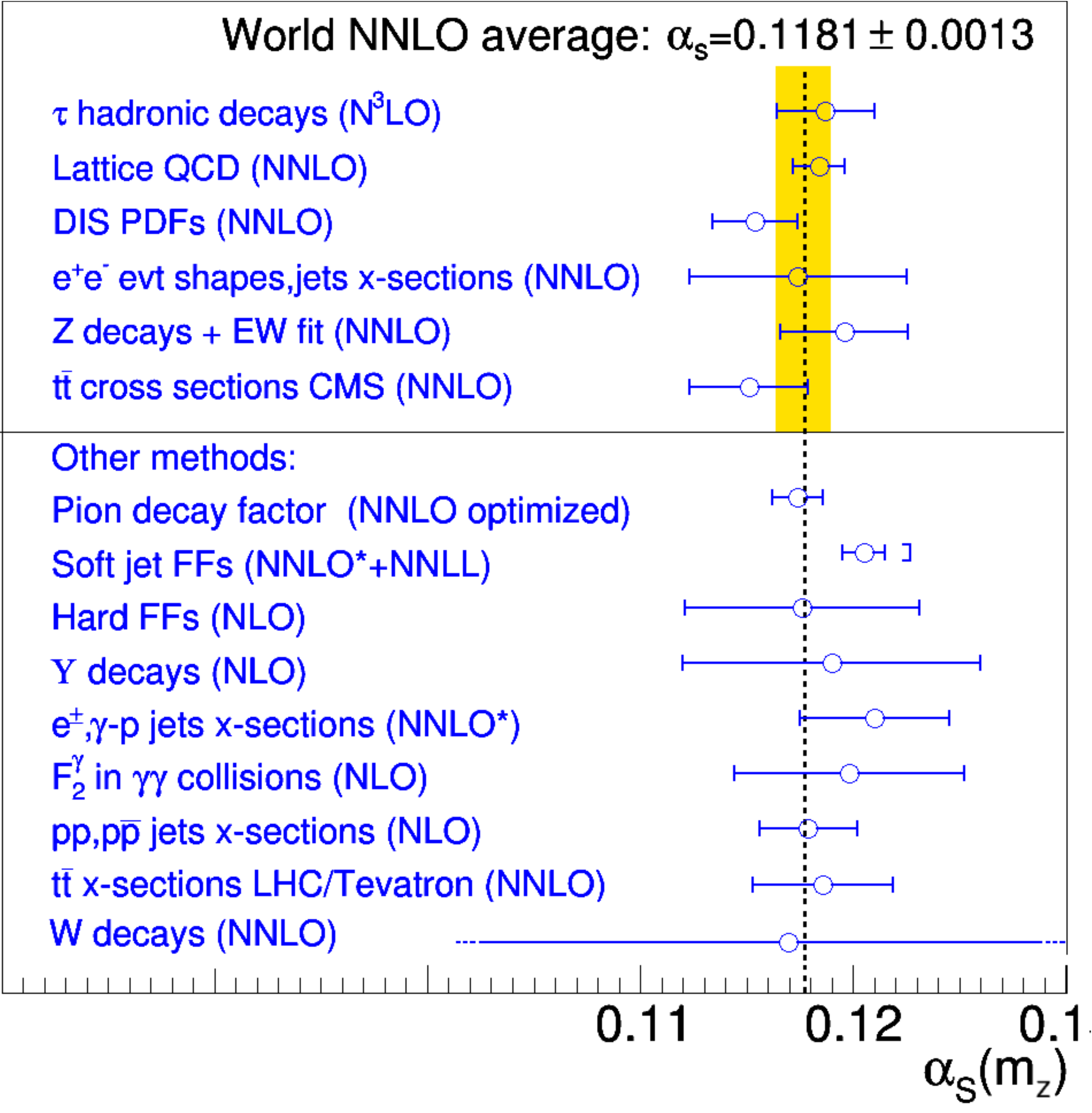}
}
\caption[]{Left: 6 classes of observables used to determine the current $\alphas$ world average (dotted line and grey band).
Dashed lines and shaded (yellow) bands indicate the pre-average values of each subclass~\cite{PDG}.
Right: Summary of all $\alphas$ extraction methods: 6 world-average classes (top), and 8 other
methods at lower level of accuracy (bottom).
}
\label{fig:alphas_averages}
\end{figure}

The six methods used in the latest global $\alphasmZ$ extraction are shown in Fig.~\ref{fig:alphas_averages} (left, and top-right)
roughly listed by increasing energy scale~\cite{PDG}:
\begin{enumerate}
\item The comparison of NNLO pQCD predictions to computational {\bf lattice QCD} ``data'' (Wilson loops, quark potentials, vacuum
polarization,..) yields $\alphasmZ = 0.1187 \pm 0.0012$, and provides the most precise $\alphas$ extraction today. 
Its $\delta\alphasmZ/\alphasmZ =$~1\% uncertainty (dominated by finite lattice spacing and statistics) has, however, 
doubled since the previous PDG pre-average due to a new calculation of the QCD static energy~\cite{Bazavov:2014soa} which is 
lower than the rest of lattice-QCD analyses. The expected improvements in computing power over the 
next 10 years would reduce the $\alphas$ uncertainty down to 0.3\%. Further reduction to the $\sim$0.1\% 
level requires the computation of 4th-order pQCD corrections.
\item The ratio of hadronic to leptonic {\bf tau decays}, known experimentally to within $\pm0.23\%$ ($R_{\rm \tau,exp} = 3.4697 \pm 0.0080$),
compared to pQCD at next-to-NNLO (N$^3$LO) accuracy, yields $\alphasmZ$~=~0.1192~$\pm$~0.0018,
\ie $\delta\alphasmZ/\alphasmZ$~=~1.5\%. This uncertainty has slightly increased (from $\pm1.3\%$) compared 
to the previous PDG revision to cover the different results obtained by various pQCD approaches (FOPT vs. CIPT, 
with different treatments of non-pQCD corrections)~\cite{Pich:2016bdg}. High-statistics $\tau$ spectral functions 
(\eg\ from B-factories, or ILC/FCC-ee in the future) and solving CIPT--FOPT discrepancies (and/or N$^4$LO calculations, 
within a $\sim$10 years time scale) are needed to bring $\alphas$ uncertainties below $\sim$1\%.
\item The QCD coupling has been obtained from various analyses of {\bf proton structure functions} (including N$^3$LO fits of 
$\rm F_{2}(x,Q^2), F^c_2(x,Q^2), F_L(x,Q^2)$, as well as global (approximately) NNLO fits of PDFs) yielding a central 
value lower than the rest of methods: $\alphasmZ$ = 0.1156~$\pm$~0.0023, with a moderate precision $\delta\alphasmZ/\alphasmZ = 2\%$
(slightly increased from the previous $\pm1.7\%$, driven by the spread of different theoretical extractions).
Resolving the differences among fits, and/or full-NNLO global fits of DIS+hadronic data (including consistent 
treatment of heavy-quark masses) would yield an $\alphas$ extraction with $\sim$1\% uncertainty. Ultimate uncertainties
in the $\delta\alphasmZ/\alphasmZ\approx$~0.15\% range require large-statistics studies at a future DIS
machine (such as LHeC or FCC-eh)~\cite{Cooper-Sarkar:2016udp}.
\item Combining the LEP data on {\bf $\epem$ event shapes and rates} (thrust, C-parameter, N-jet cross sections) with 
N$^{2,3}$LO computations (matched, in some cases, with soft and collinear resummations at N$^{(2)}$LL accuracy),
one obtains $\alphasmZ$~=~0.1169~$\pm$~0.0034. The $\delta\alphasmZ/\alphasmZ$~=~2.9\% uncertainty is mostly driven by the span 
of individual extractions which use different (Monte Carlo or more analytical) approaches to correct for hadronization effects.
Reduction of the non-pQCD uncertainties, \eg\ through new $\epem$ jet data
at lower (higher) $\sqrts$ for the event shapes (jet rates), plus jet cross sections with improved resummation (beyond NLL), 
are needed to reach $\alphas$ uncertainties below~1\%.
\item Three closely-related {\bf Z hadronic decays} observables measured at LEP ($\Rlz = \Ghad/\Gamma_\ell$, 
$\so = 12 \pi/m_Z \cdot \Gamma_e\Ghad/\Gamma_Z^2$, and $\GZ$) compared to N$^3$LO calculations, yield~\cite{Baak:2014ora} 
$\alphasmZ = 0.1196 \pm 0.0030$ with $\delta\alphasmZ/\alphasmZ\approx$~2.5\%.
Uncertainties at the permil level will require high-precision and large-statistics measurements accessible \eg with 
10$^{12}$ Z bosons at the FCC-ee~\cite{TLEP} (and associated 5-loop calculations, with reduced parametric uncertainties).
\item {\bf Top-pair cross sections}, theoretically known at NNLO+NNLL, are the first hadron
collider measurements that constrain $\alphas$ at NNLO accuracy. From the comparison of CMS data to pQCD, 
one obtains $\alphasmZ = 0.1151 \pm 0.0028$ with a $\delta\alphasmZ/\alphasmZ$~=~2.5\% uncertainty 
(mostly dominated by the gluon PDF uncertainties)~\cite{Chatrchyan:2013haa}. Preliminary combination of all $\ttbar$ 
measurements at LHC and Tevatron increases its value to $\alphasmZ = 0.1186 \pm 0.0033$.
\end{enumerate}

The $\chi^2$-average of the unweighted values for these 6 subgroups of observables (dashed lines and shaded (yellow) 
bands in Fig.~\ref{fig:alphas_averages} left) is $\alphasmZ = 0.1181~\pm~0.0013$, with a 
$\delta\alphasmZ/\alphasmZ =$~1.1\% uncertainty (dotted line and grey band in Fig.~\ref{fig:alphas_averages} left, and top-right panels)~\cite{PDG}.

\section{Other $\alphas$ extractions}
\label{sec:others}

There exist at least 8 other classes of observables, often computed at a lower accuracy (NLO, or 
approximately-NNLO, aka.\@\xspace NNLO*), used to determine the QCD coupling (Fig.~\ref{fig:alphas_averages} right, bottom), 
but not yet included in the world-average. Ordered by their energy scale, those are:

\begin{itemize}
\item The {\bf pion decay factor} ($\rm F_{\rm \pi,exp} = 92.2~\pm~0.03~\pm~0.14$~MeV) has been used to extract~\cite{Kneur:2013coa} 
$\alphasmZ$ = 0.1174~$\pm$~0.0017. Although the calculation is (``optimized'') NNLO, the low scales involved 
challenge the validity of the pQCD approach.
\item The jet-energy dependence of the {\bf soft} (low-$z$) {\bf parton-to-hadron fragmentation functions (FF)}, 
provides $\alphasmZ$ = 0.1205~$\pm$~0.0022 at NNLO*+NNLL accuracy, with a $\sim$2\% uncertainty~\cite{softFF}, 
which could be halved including full-NNLO corrections. 
\item $\gamma$-$\gamma$ measurements of the {\bf photon structure function} F$_2^\gamma(x,Q^2)$ have been used to obtain
$\alphasmZ$ = 0.1198~$\pm$~0.0054 at NLO~\cite{Albino:2002ck}, with $\delta\alphasmZ/\alphasmZ\approx$~4.5\%.
Extension to NNLO (and inclusion of new B-factories data) would reduce this uncertainty to $\sim$2\%.
\item The {\bf $\Upsilon$ decay} ratio $\rm R_\gamma\equiv \Gamma(\Upsilon(1S) \to \gamma\, X)/\Gamma(\Upsilon(1S) \to X)$ 
(with X = light hadrons) has been computed at NLO accuracy in the NRQCD framework.
From the CLEO data one obtains $\alphasmZ$ = 0.119~$\pm$~0.007, with a $\sim$6\%,
uncertainty shared equally by experimental and theoretical systematics~\cite{Brambilla:2007cz}. NNLO corrections with improved 
long-distance matrix elements, 
and more precise measurements of the $\gamma$ spectrum 
(and of the parton-to-photon FF) would allow for an extraction with $\delta\alphasmZ/\alphasmZ\approx$~2\% in a few years from now.
\item From the scaling violations of the {\bf hard} (high-$z$) {\bf parton-to-hadron FFs} one extracts
$\alphasmZ$ = 0.1176~$\pm$~0.0055 at NLO, with $\sim$5\% uncertainties, mostly of experimental origin~\cite{Albino:2005me}. 
Extension of the global FF fits at NNLO accuracy, and inclusion of new datasets (already available at B-factories) 
would allow reaching $\delta\alphasmZ/\alphasmZ\approx$~2\%.
\item The NNLO$^*$ calculation of {\bf jet cross sections in DIS and photoproduction} provides $\alphasmZ$ = 0.120~$\pm$~0.004  with 
$\delta\alphasmZ/\alphasmZ\approx$~3\% precision today~\cite{Biekotter:2015nra}. Upcoming full-NNLO analyses~\cite{Currie:2016ytq}
could reduce this uncertainty to the $\sim$1.5\% level, whereas a future DIS machine (such as LHeC or FCC-eh) would further bring it below 1\%.
\item Measurements of {\bf W hadronic decays}, although computed at N$^3$LO, provide today a very 
imprecise $\alphasmZ$ = 0.117~$\pm$~0.030 with $\pm$25\% uncertainty, due to the poor LEP data~\cite{d'Enterria:2016ujp}. 
A competitive $\alphas$ extraction requires statistical samples of 10$^8$ W, available at FCC-ee, 
which (combined with N$^4$LO corrections) can ultimately yield $\delta\alphasmZ/\alphasmZ\approx$~0.1\%.
\item Various {\bf jet observables at hadron colliders} (ratio of 3- to 2-jets, 3-jet mass, inclusive cross sections)
have tested asymptotic freedom at TeV scales. Combining those, one obtains $\alphasmZ$ = 0.1179~$\pm$~0.0023 
at NLO accuracy, with $\delta\alphasmZ/\alphasmZ\approx$~2\% dominated by theoretical uncertainties. 
The imminent incorporation of NNLO corrections~\cite{Currie:2013dwa} and a consistent combination 
(including correlations) of the multiple datasets available at Tevatron and LHC, may
reduce the $\alphas$ uncertainties to the 1.5\% level in the upcoming years.
\end{itemize}

Assuming all 14 extraction methods discussed here are computed at NNLO (or above) accuracy, and 
{\em provided that} they yield consistent $\alphas$ results, a simple weighted-average would have an uncertainty of
$\delta\alphasmZ/\alphasmZ \approx$~0.35\%, $\sim$3 times better than the present value. 
A permil-level $\alphas$ uncertainty requires high-precision 
future $\epem$ colliders with very large Z and W samples, complemented with 
4$^{\rm th}$-order pQCD corrections, and improved parametric uncertainties.\\

\noindent {\bf Acknowledgments} I am grateful to S.~Bethke and G.~Salam for useful discussions, and to R.~P\'erez-Ramos 
and M.~Srebre for common work in two of the new $\alphas$ extractions reported~here.

\end{document}